\documentclass[letterpaper,12pt]{article}

\usepackage{times}
\usepackage[margin=1in]{geometry}

\usepackage{graphicx}
\DeclareGraphicsExtensions{.pdf,.jpeg,.png}
\usepackage{cite}
\usepackage{amsmath}
\usepackage{amsfonts}
\usepackage{titlesec}
\usepackage{enumerate}
\usepackage[linesnumbered,ruled]{algorithm2e}
\usepackage[justification=centering]{caption}
\titlelabel{\thetitle.\quad}
\titleformat*{\section}{\normalfont\bfseries}
\titleformat*{\subsection}{\normalfont\bfseries}
\titleformat*{\subsubsection}{\normalfont\bfseries}
\titleformat*{\paragraph}{\normalfont\bfseries}
\titleformat*{\subparagraph}{\normalfont\bfseries}

\usepackage{hyperref}
\hypersetup{hypertex=true,
	colorlinks=true,
	linkcolor=red,
	anchorcolor=blue,
	citecolor=green}
\usepackage[all]{xy}
\usepackage{arydshln}

\begin{document}
\date{}

\title{Kalman Filter from the Mutual Information Perspective}

\author{Yarong Luo, 
	yarongluo@whu.edu.cn\\
	Jianlang Hu,
	hujianlang123@whu.edu.cn\\
 	Chi Guo,
	guochi@whu.edu.cn\\
 	GNSS Research Center, Wuhan University
}



\maketitle

\thispagestyle{empty}

\noindent
{\bf\normalsize Abstract}\newline
{Kalman filter is a best linear unbiased state estimator. It is also comprehensible from the point view of the Bayesian estimation. However, this note gives a detailed derivation of Kalman filter from the mutual information perspective for the first time. Then we extend this result to the R\'enyi mutual information. Finally we draw the conclusion that the measurement update of the Kalman filter is the key step to minimize the uncertainty of the state of the dynamical system. 
} \vspace{2ex}

\noindent
{\bf\normalsize Key Words}\newline
{Kalman filter, mutual information,  R\'enyi mutual information,  uncertainty, measurement update}

\section{Introduction}
Kalman filter has been widely in various fields as an effective state estimator for integrated navigation~\cite{gongmin2017lectures}, robotics~\cite{thrun2005probabilistic}, etc. The classical Kalman filter can be derived as a best linear unbiased estimate~\cite{gongmin2017lectures} and it is easy understand it from the probabilistic perspective~\cite{thrun2005probabilistic}. 
Recently, Kalman filter has also been presented using the methods of maximum relative entropy~\cite{giffin2014kalman} and the temporal derivative of the R\'enyi entropy~\cite{luo2020novel}, which go beyond the general Bayesian filter. 
More and more evidences show that Kalman filter can be regarded as a direct extension of information theory. 
This note gives a new perspective of the Kalman filter from the mutual information, which further bridges the gap between the optimal state estimation and the information theory.
The main contribution of this note is to derive the Kalman filter from the perspective of mutual information and extend it to the R\'enyi mutual information case. 
\section{Kalman Filter from the Mutual Information}
For following discrete-time state-space model
\begin{equation}
\label{discrete_KF1}
X_k={\Phi_{k|k-1}}X_{k-1}+ {\Gamma_{k|k-1}}W_{k-1}
\end{equation}
\begin{equation}
\label{discrete_KF2}
Z_k=H_kX_k+V_k
\end{equation}
where $X_k$ is n-dimensional state vector;
$Z_k$ is m-dimensional measurement vector;
${\Phi_{k|k-1}}$, ${\Gamma_{k|k-1}}$ and $H_k$ are the known system structure parameters, which are called the $n\times n$ dimensional one-step state update matrix, the $n \times l$ dimensional system noise distribution matrix, and the $m \times n$ dimensional measurement matrix, respectively; 
$W_{k-1}$ is the $l$-dimensional system noise vector, and $V_k$ is the m-dimensional measurement noise vectors.
Both of them are Gaussian noise vector sequences with zero mean value, and independent to each other:
\begin{equation}
\label{discrete_noise_1}
\mathbb{E}[W_k]	=0,\mathbb{E}[W_k W_j^T]	=Q_k \delta_{kj}
\end{equation}
\begin{equation}
\label{discrete_noise_2}
\mathbb{E}[V_k]	=0,\mathbb{E}[V_k V_j^T]	=R_k \delta_{kj}
\end{equation}
\begin{equation}
\label{discrete_noise_3}
\mathbb{E}[W_k V_j^T]	=0
\end{equation}

The one-step prediction covariance matrix is denoted as $\Sigma_{k|k-1}$. The state estimation at $t_k$ is denoted as $\mathcal{N}(\hat{X}_k,\Sigma_{k})$, where $\hat{X}_k$ is the mean of estimated state and $\Sigma_{k}$ is the covariance matrix of the estimated covariance matrix. 
Assuming the optimal estimation of the state can be calculated as follows:
 \begin{equation}
 \label{estimation_correction1}
 \hat{X}_k={X_{k|k-1}^-}+K_k{\tilde{Z}_{k|k-1}}
 \end{equation}
 where $K_k$ is the undetermined correction factor matrix, $X_{k|k-1}^-=X_{k|k-1}-\hat{X}_{k-1}$ is the state estimation error, $\tilde{Z}_{k|k-1}=Z_k-H_kX_{k|k-1}^-$ is the measurement one-step prediction error.

Then, the mean square error matrix of state estimation $\hat{X}_k$ is given by~\cite{gongmin2017lectures}
\begin{equation}
\label{estimation_correction5}
{\Sigma}_k=({I}-K_kH_k){{\Sigma}_{k|k-1}}({I}-K_kH_k)^T+K_kR_kK_k^T
\end{equation}
the mean square error matrix $\Sigma_{k}$ is positive definite as $({I}-K_kH_k){{\Sigma}_{k|k-1}}({I}-K_kH_k)^T$ is positive definite and $K_kR_kK_k^T$ is positive definite.

The joint Gaussian distribution can be expressed
\begin{equation}\label{joint_PDF}
p(X,Y)\sim \left(\begin{bmatrix}
\hat{X}\\ \hat{Y}
\end{bmatrix},\begin{bmatrix}
\Sigma_{xx} &\Sigma_{xy}\\ \Sigma_{yx} & \Sigma_{yy}
\end{bmatrix} \right)
\end{equation}
where $X\sim \mathcal{N}(\hat{X},\Sigma_{xx})$,$Y\sim \mathcal{N}(\hat{Y},\Sigma_{yy})$.

The mutual information for a joint Gaussian PDF can be represented by
\begin{equation}
\label{MI_1_1}
\begin{aligned}
I(X,Y)&=H(X)+H(Y)-H(X,Y)=H(X)-H(X|Y)\\
&=\frac {1}{2}\ln ((2\pi e)^N \det \Sigma_{xx})+\frac {1}{2}\ln ((2\pi e)^M \det \Sigma_{yy})-\frac {1}{2}\ln ((2\pi e)^{M+N} \det \Sigma)\\
&=-\frac {1}{2}\ln \left(\frac{\det \Sigma}{\det\Sigma_{xx} \det \Sigma_{yy}}\right)
\end{aligned}
\end{equation}
 where $H(X)=\frac {1}{2}\ln ((2\pi e)^N \det \Sigma_{xx})$ is the entropy of a Gaussian random variable, and 
\begin{equation}
\label{matrix_equality}
\det \Sigma=\det \Sigma_{xx}\det\left(\Sigma_{yy}-\Sigma_{yx}\Sigma_{xx}^{-1}\Sigma_{xy}\right)
=\det \Sigma_{yy}\det \left(\Sigma_{xx}-\Sigma_{xy}\Sigma_{yy}^{-1}\Sigma_{yx}\right)
\end{equation}

Therefore, the mutual information describes the reduction of the uncertainty in variable X due to gaining knowledge of variable Y.

Similarly, the mutual information at time step $t_{k+1}$ can be easily computed by the a priori, a posteriori PDF and the Kalman gain $K_k$ as
\begin{equation}\label{MI_KF}
I(\hat{X}_{k|k-1},Z_k)=\frac{1}{2}\ln \left( \frac{\det \Sigma_{k|k-1}}{\det \Sigma_{k} } \right)
=\frac{1}{2}\ln \left( \frac{\det \Sigma_{k|k-1}}{\det \left(({I}-K_kH_k){{\Sigma}_{k|k-1}}({I}-K_kH_k)^T+K_kR_kK_k^T\right) } \right)
\end{equation}

It describes the reduction of the uncertainty of the state due to gaining knowledge from the measurement $Z_k$. Consequently, we want to maximize the mutual information $I(\hat{X}_{k|k-1},Z_k)$. It is obvious that the maximum of the mutual information is equivalent to the minimum of $\ln\det \Sigma_{k}$.
We can note that equation (\ref{MI_KF}) is the function of the unknown factor matrix and thereby the maximization of it can be calculated by taking derivative of it with respect to $K_k$ and setting it equal to zero,
\begin{equation}\label{derivative_K_k}
\frac{d I(\hat{X}_{k|k-1},Z_k)}{d K_k}=-\Sigma_{k}^{-T}\frac{d\Sigma_{k}}{dK_k}=-\Sigma_{k}^{-T}\left(-2(I-K_kH_k)\Sigma_{k|k-1}H_k^T+2K_kR_k  \right)=0
\end{equation}
where $\frac{\partial \ln\det X}{\partial X}=X^{-T}$~\cite{zhang2017matrix} has been used and then solving for $K_k$ gives
\begin{equation}\label{Kalman_filter_gain}
K_k=\Sigma_{k|k-1}H_k^T(H_k\Sigma_{k|k-1}H_k^T+R_k)^{-1}
\end{equation}

A subsequent derivative of equation (\ref{derivative_K_k}) must be performed to check for a maximum, that is 
\begin{equation}\label{maximum}
\frac{d}{dK_k}\left(-\Sigma_{k}^{-T}\left(-2(I-K_kH_k)\Sigma_{k|k-1}H_k^T+2K_kR_k  \right)\right)
\end{equation}

After substituting the equation (\ref{derivative_K_k}) into above equation results in
\begin{equation}\label{two_maximu}
\frac{d}{dK_k}\left(-\Sigma_{k}^{-T}\left(-2(I-K_kH_k)\Sigma_{k|k-1}H_k^T+2K_kR_k  \right)\right)=-\Sigma_{k}^{-T}(H_k\Sigma_{k|k-1}H_k^T+R_k)
\end{equation}
which is always negative definite by the definition of the covariance matrices $R_k$ and $H_k\Sigma_{k|k-1}H_k^T$, ensuring the solution for $K_k$ is a maximum.
\section{Kalman Filter from the R\'enyi Mutual Information}
 Moreover, the R\'enyi mutual information of a joint Gaussian PDF can be calculated similar to equation (\ref{MI_1_1}):
 \begin{equation}
 \label{Renyi_joint_pdf}
 \begin{aligned}
 I_R^{\alpha}(X,Y)&=H_R^{\alpha}(X)+H_R^{\alpha}(Y)-H_R^{\alpha}(X,Y)\\
 &=\frac{1}{2}\ln |(2\pi)^{N} \alpha^{\frac{N}{\alpha-1}} \det \Sigma_{xx}|+\frac{1}{2}\ln |(2\pi)^{M} \alpha^{\frac{M}{\alpha-1}} \det \Sigma_{yy}|-\frac{1}{2}\ln |(2\pi)^{N+M} \alpha^{\frac{N+M}{\alpha-1}} \det \Sigma|\\
 &=\frac {1}{2}\ln \left(\frac{\det\Sigma_{xx} \det \Sigma_{yy}}{\det \Sigma}\right)\\
 &=I(X,Y)
 \end{aligned}
 \end{equation}
 where $H_R^{\alpha}(X)=\frac{1}{2}\ln |(2\pi)^{N} \alpha^{\frac{N}{\alpha-1}} \det \Sigma_{xx}|$ is the R\'enyi entropy of order $\alpha$ for a continuous Gaussian random variable with a multivariate Gaussian PDF.
 Consequently, we can know that the mutual information is the same as the R\'enyi mutual information for the joint Gaussian PDF.
 Similarly, we can get the same result as equation (\ref{Kalman_filter_gain}) from the R\'enyi mutual information.
\section{Conclusions}
In this paper, Kalman filter is derived from the perspective of mutual information and extended to the R\'enyi mutual information case. 
We show that the measurement update of the Kalman filter can minimize the uncertainty of the state by formulating it as the mutual information between the evolving state and the measurement and maximizing the mutual information. 
Furthermore, we can think of Kalman filter a little more radically as an extension of the information theory.  

\vspace{2ex}

\noindent
{\bf\normalsize Acknowledgement}\newline
{This research was supported by a grant from the National Key Research and Development Program of China (2018YFB1305001). 
We express thanks to GNSS Center, Wuhan University.} \vspace{2ex}

\bibliographystyle{IEEEtran}
\bibliography{ref.bib}

\end{document}